\documentclass[12pt]{article}

\ifx\pdfoutput\undefined
\usepackage[dvips,bookmarks]{hyperref}
\else
\usepackage{hyperref}
\fi
\hypersetup{colorlinks=false,bookmarksopen,bookmarksnumbered,citecolor=blue,
   pdfstartview=FitH}

\usepackage[dvips]{graphicx}
\usepackage{latexsym}

\usepackage{amssymb,amsfonts,amsmath}
\usepackage{graphicx} 
\usepackage{indentfirst}

 \usepackage{bbm}

\parskip=\medskipamount

\newcommand {\cC}{{\cal C}}
\newcommand {\cD}{{\cal D}}
\newcommand {\cE}{{\cal E}}

\newcommand {\cH}{{\cal H}}

\newcommand {\cJ}{{\cal J}}
\newcommand {\cK}{{\cal K}}
\newcommand {\cL}{{\cal L}}
\newcommand {\cM}{{\cal M}}
\newcommand {\cN}{{\cal N}}

\newcommand {\cR}{{\cal R}}
\newcommand {\cS}{{\cal S}}

\newcommand {\cV}{{\cal V}}
\newcommand {\cW}{{\cal W}}

\def\a{\alpha}

\def\b{\beta}

\def\d{\delta}
\def\e{\epsilon}

\def\g{\gamma}
\def\G{\Gamma}

\def\l{\lambda}
\def\m{\mu}
\def\n{\nu}
\def\o{\omega}

\def\q{\theta}
\def\r{\rho}
\def\s{\sigma}
\def\t{\tau}

\def\D{\Delta}
\def\F{\Phi}
\def\J{\Psi}
\def\L{\Lambda}
\def\O{\Omega}

\def\S{\Sigma}
\def\U{\Upsilon}
\def\X{\Xi}
\def\tr{{\rm tr}}
\def\rd{{\rm d}}
\def\ri{{\rm i}}
\def\re{{\rm e}}

\newcommand{\ve}{\varepsilon}                            
\newcommand{\cDB}{{\bar\cD}}                            

\newcommand{\ab}{{\a\b}}

\newcommand{\pa}{\partial}                           
\newcommand{\hf}{\frac12}

\newcommand{\vf}{\varphi}

\newcommand{\be}{\begin{equation}}
\newcommand{\ee}{\end{equation}}
\newcommand{\bea}{\begin{eqnarray}}
\newcommand{\eea}{\end{eqnarray}}
\newcommand{\non}{\nonumber}
\newcommand{\bm}[1]{\mbox{\boldmath$#1$}}

\def\double #1{#1{\hbox{\kern-2pt $#1$}}}

\newcommand{\alu}{{\underline{\a}}}
\newcommand{\beu}{{\underline{\b}}}
\newcommand{\gau}{{\underline{\g}}}

\newcommand{\teb}{{\bar{\theta}}}
\newcommand{\qb}{{\bar{\theta}}}

\newcommand{\UE}{{\underline{E}}}

\newif\ifdtup

\newcommand{\bsubeq}{\begin{subequations}}
\newcommand{\esubeq}{\end{subequations}}

\numberwithin{equation}{section}

\begin{document}
\begin{titlepage}
\begin{flushright}
September, 2012\\
\end{flushright}
\vspace{5mm}

\begin{center}
{\Large \bf 
Prepotentials for $\bm{\cN=2}$ conformal  supergravity \\
 in three dimensions}\\ 
\end{center}

\begin{center}

{\bf
 Sergei M. Kuzenko
} \\
\vspace{5mm}

\footnotesize{
{\it School of Physics M013, The University of Western Australia\\
35 Stirling Highway, Crawley W.A. 6009, Australia}}  
\\
\vspace{2mm}

\end{center}

\begin{abstract}
\baselineskip=14pt
We present a complete solution of the constraints for three-dimensional 
$\cN=2$ conformal supergravity in terms of unconstrained prepotentials. 
This allows us to develop a prepotential description of the off-shell 
versions of $\cN=2$ Poincar\'e and anti-de Sitter supergravity theories 
constructed in arXiv:1109.0496. 
\end{abstract}

\vfill

\vfill
\end{titlepage}

\newpage
\renewcommand{\thefootnote}{\arabic{footnote}}
\setcounter{footnote}{0}

%


\section{Introduction}
\setcounter{equation}{0}

$\cN=2$ conformal   supergravity in three dimensions
\cite{RvN} has recently been formulated in an off-shell superspace setting 
\cite{KLT-M11}, building on the conventional torsion constraints given in \cite{HIPT}.
This formulation has then been employed in \cite{KT-M11} to construct several off-shell 
versions for $\cN=2$ Poincar\'e and anti-de Sitter supergravity theories 
by coupling the Weyl supermultiplet to certain conformal compensators. 
Here we provide a complete solution of the constraints for three-dimensional 
$\cN=2$ conformal supergravity in terms of unconstrained prepotentials.
In particular, we show that, modulo gauge transformations, the Weyl supermultiplet 
is described by a real vector superfield $H_{\ab} = H_{\b\a} $ with a nonlinear gauge 
transformation law 
\bea
\d H_{\a\b} = \bar D_{(\a} L_{\b)} - D_{(\a} \bar L_{\b)} +O(H)~, 
\label{1.1}
\eea
where the gauge parameter $L_\a$ is an unconstrained complex spinor. 
The linearized version of this transformation law was given in \cite{KT-M11}.
The prepotential $H_{\a\b}$ is a three-dimensional (3D) analog of the 
$\cN=1$ gravitational superfield  in four dimensions introduced for the first time 
in \cite{OS1,OS2}.\footnote{See \cite{BK} for a review
of the Ogievetsky-Sokatchev approach to $\cN=1$ 
supergravity in four dimensions.}

Our three-dimensional construction is similar
to the famous  prepotential description,
originally due to Siegel \cite{Siegel} (and further developed in \cite{SG,GS80}),
for the Wess-Zumino  supergravity formulation
\cite{WZ} (also known, in the component approach, 
as the old minimal formulation\footnote{The linearized version of old minimal supergravity 
appeared in  \cite{OS1,FZ}.}
for 4D $\cN=1$ supergravity \cite{old}), 
and its extension \cite{GGRS}  to the case of  4D $\cN=1$ 
conformal supergravity realized in superspace \cite{Howe}.
At the same time, as will be shown below, some nontrivial differences exist between  
the 3D $\cN=2$ and 4D $\cN=1$ supergravity theories. In principle, 
the 3D prepotential solution could be obtained by dimensional reduction from 
4D $\cN=1$ conformal supergravity following the procedure outlined in section 
7.9 of {\it Superspace} \cite{GGRS}. In practice, however, it is more advantageous to follow 
a manifestly covariant approach and derive the solution from scratch. 
In this sense the 3D story  is similar to that of (2,2) supergravity in two dimensions
\cite{GrisaruW1,GrisaruW2}.

For both $\cN=2$ Poincar\'e and anti-de Sitter supergravity theories  in three dimensions,
the linearized off-shell actions and associated supercurrent multiplets\footnote{See 
\cite{Dumitrescu:2011iu} for an alternative construction of 3D $\cN=2$ supercurrents.} 
were constructed in \cite{KT-M11} by applying dimensional reduction and duality. 
Using the prepotential formulation developed in this paper, 
these results can be re-derived from first principles. 
In four dimensions, the supercurrent \cite{FZ0}
originates as the source of supergravity  \cite{OS1,FZ,Siegel0}. 
Now we have the same picture in the 3D case.

The prepotential formulation given in this paper makes 
it possible to develop supergraph techniques 
for 3D $\cN=2$ matter-coupled supergravity,  including a three-dimensional extension of 
the background-field formalism for $\cN=1$ supergravity in four dimensions \cite{GS1,GS2}.
Such techniques may be useful to address several essentially three-dimensional problems. 
First of all, the supergraph techniques may be useful to study 
the quantum properties of new massive supergravity theories
\cite{Andringa:2009yc,Bergshoeff:2010mf,Bergshoeff:2010ui}.
Secondly, they  may be useful for explicit calculations of newly 
discovered anomalies in three dimensions  \cite{CDFKS1,CDFKS2}. 
Thirdly, they may be helpful for achieving  a better understanding of quantum (super)gravity 
in three dimensions.
As is known, 3D Poincar\'e and anti-de Sitter (super)gravity theories do not have
propagating degrees of freedom at the classical level, since there are no gravitational 
waves in three dimensions. At the quantum level, 3D gravity can be consistently defined, 
within a Chern-Simons formulation, and constitutes  an exactly soluble model  \cite{Witten}, 
in spite of earlier beliefs that general relativity in three dimensions was unrenormalizable. 
The same conclusions hold for $\cN$-extended supergravity theories in three dimensions
formulated as  Chern-Simons theories in \cite{AT,AT2}. 
It would be interesting to see how the exact solubility of 3D (super)gravity 
manifests  itself in terms of the standard effective action derived using conventional geometric formulations. The case of 3D $\cN=2$ supergravity formulated in superspace provides 
a simple playground for that.  

This paper is organized as follows. In section 2 we briefly review the superspace formulation 
for $\cN=2$ conformal supergravity given in \cite{KLT-M11}. We also present 
a $\cN=2$ supersymmetric generalization of the Cotton tensor.
The prepotential formulation for $\cN=2$ conformal supergravity is described in 
sections three to five. Finally,  a discussion of the results is given in section 5.
We use the 3D conventions of \cite{KLT-M11}.

\section{Conformal supergravity}

In this section we give a brief review of the superspace formulation 
for $\cN=2$ conformal supergravity presented in \cite{KLT-M11}. 
It makes use of a curved 3D $\cN=2$ superspace  $\cM^{3|4}$ parametrized by
local bosonic ($x$) and fermionic ($\q, \bar \q$)
coordinates  
$$z^{{M}}=(x^{m},\q^{\mu},{\bar \q}_{{\mu}}),\qquad
m=0,1,2~, \qquad\mu=1,2~,
$$
where the Grassmann variables $\q^{\mu} $ and $\teb^{\mu} = \ve^{\m\n} \teb_{\nu}$
are mutually conjugate. In accordance with \cite{HIPT}, 
the structure group is chosen to be ${\rm SL}(2,{\mathbb{R}})\times {\rm U(1)}_R$.
The covariant derivatives have the form
\bea
\cD_{{A}} =(\cD_{{a}}, \cD_{{\a}},\cDB^\a) =E_A +\O_A{}^{\b\g}\cM_{\b\g} 
+\ri \,\F_{{A}}\cJ~,
\label{CovDev}
\eea
where $E_{{A}}=E_{A}{}^{M}(z) \pa/\pa z^{{M}}$
is the supervielbein, $\O_A{}^{\b\g}$ and $\F_A$ are the Lorentz and the U$(1)_R$ connections
respectively.\footnote{Our normalization of the Lorentz connection and the Lorentz curvature 
differs by a factor of 1/2 from that adopted in  \cite{KLT-M11,KT-M11}.}  
The Lorentz generators act on the covariant derivatives as follows:
\bea
{[}\cM_{\g\d},\cD_{a}{]}
= \ve_{abc} (\g^c)_{\g \d} \cD^b~, \qquad
{[}\cM_{\g\d},\cD_{\a}{]}
=\ve_{\a(\g}\cD_{\d)}~. 
\eea
This can be rewritten as 
\bea
[\L^{\g\d} \cM_{\g\d} , \cD_A] = \L_A{}^B\cD_B 
=  \left\{\begin{array}{c}
\phantom{-}\L_a{}^b \cD_b  \\
\phantom{-}\L_\a{}^\b \cD_\b \\
-\L^\a{}_\b \bar \cD^\b 
\end{array}
\right. ~,
\eea
where we have defined 
\bea
\L_{ab} :=  \ve_{abc} (\g^c)_{\g \d} \L^{\g\d} ~.
\eea
The U$(1)_R$ generator is defined to act on the covariant derivatives by the rule:
\bea
[\cJ, \cD_A] = q_A \cD_A \quad \longleftrightarrow \quad 
[\cJ, \cD_a]=0~, \quad
[\cJ , \cD_\a ] = \cD_\a~, \quad [\cJ , \bar \cD^\a ] =-\bar \cD^\a~.
\eea
The supergravity gauge group is described by local transformations
of the form
\be
 \cD_{{A}} \to \re^\cK \cD_A \re^{-\cK}~,
\qquad \cK = K^{{C}}(z) D_{{C}} +l^{ \g\d }(z) \cM_{\g\d}
+\ri \, \t (z) \cJ  ~,
\label{tau}
\ee
with $D_M = (\pa_m , D_\m , \bar D^\m)$ the flat superspace covariant derivatives.
The gauge parameters in \eqref{tau}
obey natural reality conditions, but otherwise  are arbitrary.
Given a tensor superfield $U(z)$, with its indices suppressed,
its transformation law is
\bea
 U \to \re^\cK U~.
\label{tensor-K}
\eea

The  torsion and curvature tensor are defined by 
\bea
{[}\cD_{{A}},\cD_{{B}}\}&=&
T_{ {A}{B} }{}^{{C}}\cD_{{C}}
+ R_{{A} {B}}{}^{{\g\d}}\cM_{\g\d}
+\ri \,F_{ {A} {B}}\cJ
~,
\label{algebra}
\eea
where $T_{{A} {B} }{}^{{C}}$ is the torsion,
and $R_{ {A} {B}}{}^{\g\d}$  and  $R_{{A} {B}}$
are the Lorentz and U$(1)_R$ curvature tensors respectively.
Using the anholonomy coefficients 
\bea
[E_A, E_B\} = C_{AB}{}^C E_C~,
\eea
one may obtain the following explicit expressions:
\begin{subequations}\label{2.4}
\bea
T_{AB}{}^C&=& C_{AB}{}^C + \O_{AB}{}^C -(-1)^{\e_A\e_B} \O_{BA}{}^C
+\ri q_B \F_A \d_B{}^C -(-1)^{\e_A\e_B} \ri q_A \F_B \d_A{}^C ~, \label{2.4a}\\
R_{AB}{}^{\g\d}&=& E_A \O_B{}^{\g\d} -(-1)^{\e_A\e_B}E_B \O_A{}^{\g\d}  
- C_{AB}{}^C \O_C{}^{\g \d} +\O_A{}^{\g \l} \O_{B\l}{}^\d + \O_A{}^{\d \l} \O_{B\l}{}^\g~,~~~
\label{2.4b}\\
F_{AB}&=& E_A\F_B - (-1)^{\e_A\e_B} E_B\F_A - C_{AB}{}^C\F_C~. \label{2.4c}
\eea
\end{subequations}
Making use of \eqref{2.4a}, the expression for $F_{AB}$ can be rewritten as
\bea
F_{AB} = \cD_A \F_B - (-1)^{\e_A\e_B} \cD_B \F_A - T_{AB}{}^C \F_C~.
\eea

Unlike the four-dimensional case, the spinor derivatives $\cD_\a$ and $\bar \cD_\a$ 
transform in the same representation of ${\rm SL}(2,{\mathbb R})$. For practical manipulations
with objects like $T_{AB}{}^C$, it is useful to introduce a formal difference between 
the spinor indices carried by  $\cD_\a$ and $\bar \cD_\a$. Specifically, if a confusion is possible, 
we will use the following notation 
$$
\cD_{{A}} =(\cD_{{a}}, \cD_{{\a}},\cD^\alu) \equiv (\cD_{{a}}, \cD_{{\a}},\cDB^\a)~.
$$

In order to describe conformal supergravity, the torsion tensor should obey certain 
algebraic constraints \cite{HIPT}.
Imposing these constraints and then solving the Bianchi identities \cite{KLT-M11}
leads to the  algebra of covariant derivatives
\bsubeq\label{N=2-alg}
\bea
\{\cD_\a,\cD_\b\}
&=&-4\bar{\cR}\cM_{\a\b}
~,~~~~~~
\{\cDB_\a,\cDB_\b\}
= 4{\cR}\cM_{\a\b}~,
\label{N=2-alg-a}
\\
\{\cD_\a,\cDB_\b\}
&=&
-2\ri\cD_{\a\b}
-2\cC_{\a\b}\cJ
-4\ri\ve_{\a\b} \cS\cJ
+4\ri\cS\cM_{\a\b}
-2\ve_{\a\b}\cC^{\g\d}\cM_{\g\d}
~,
\label{N=2-alg-b}\\
{[}\cD_{\a\b},\cD_\g{]}
&=&
-\ri\ve_{\g(\a}\cC_{\b)\d}\cD^{\d}
+\ri\cC_{\g(\a}\cD_{\b)}
-2\ve_{\g(\a}\cS\cD_{\b)}
-2\ri\ve_{\g(\a}\bar{\cR}\cDB_{\b)}
\non\\
&&
+2\ve_{\g(\a}\cC_{\b)\d\r}\cM^{\d\r}
-\frac{4}{3}\Big(
2\cD_{(\a}\cS
+\ri\cDB_{(\a}\bar{\cR}
\Big)\cM_{\b)\g}
+\frac{1}{3}\Big(
2\cD_{\g}\cS
+\ri\cDB_{\g}\bar{\cR}
\Big)\cM_{\a\b}
\non\\
&&
+\Big(\cC_{\a\b\g}
+\frac{1}{3}\ve_{\g(\a}\Big(
8(\cD_{\b)}\cS)
+\ri\cDB_{\b)}\bar{R}
\Big)
\Big)\cJ
~,
\label{N=2-alg-c}
\eea
\esubeq
where $\cC_{\a\b\g}:= 
-\ri \cD_{(\a}\cC_{\b\g)}$.
All components of the torsion and curvature are determined in terms 
of the  three dimension-1  superfields:
a real scalar $\cS$, a
complex scalar $\cR$ and its conjugate $\bar{\cR}$,  and a real vector $\cC_a$.\footnote{Our 3D notation and conventions coincide with those used in \cite{KLT-M11}. In particular, 
given a three-vector $V_a$, it can equivalently be realized as a symmetric second-rank spinor 
$V_{\a\b} =V_{\b \a}$.
The relationship between $V_a$ and $V_{\a \b}$ is as follows:
$V_{\a\b}:=(\g^a)_{\a\b}V_a=V_{\b\a}$ and 
$V_a=-\hf(\g_a)^{\a\b}V_{\a\b}$.}
The ${\rm U(1)}_R$ charges of $\cR$ and $\bar \cR$ are
\bea
[\cJ ,\cR] =-2 \cR~, \qquad [\cJ, \bar \cR] = 2 \bar \cR~.
\eea
The torsion superfields obey differential constraints implied by the Bianchi identities, 
which are
\begin{subequations}
\bea
\cDB_\a \cR&=&0~,\\
(\cD^2
-4\bar{\cR})\cS
&=&(\cDB^2
-4\cR)\cS
=0~,  \label{2.12b}\\
\cD_{\a}\cC_{\b\g}
&=&
\ri \cC_{\a\b\g}
-\frac{1 }{ 3}\ve_{\a(\b}\Big(
 \cDB_{\g)}\bar{\cR}
+4\ri \cD_{\g)}{\cS}
\Big)~. 
\eea
\end{subequations}
Eq. \eqref{2.12b}
 means that $\cS$ is a covariantly linear superfield.
 
The contractions $\cD^2$ and $\bar \cD^2 $ in \eqref{2.12b}
are defined as
\bea
 \cD^2 := \cD^\a \cD_\a~, \qquad   \bar \cD^2 := \bar \cD_\a \bar \cD^\a~.
 \eea
 The rational for these definitions is that the operation of complex conjugation gives 
 $\overline {\cD^2 U} = \bar \cD^2  \bar U  $,  for any tensor superfield $U$.\footnote{Had we  decided
 to use one and only one type of index contraction,  for instance $\cD^2 := \cD^\a \cD_\a$ and  
 $ \bar \cD^2 := \bar \cD^\a \bar \cD_\a$, there would have appeared 
 numerous sign factors.}

The reason why the curved superspace geometry \eqref{N=2-alg}
describes $\cN=2$ conformal supergravity is that this geometry is compatible 
with super-Weyl invariance \cite{KLT-M11,HIPT}.
The super-Weyl transformation of the covariant derivatives \cite{KT-M11} is
\bsubeq\label{sW-general}
\bea
\cD{}_\a~&\to&~\re^{\hf\s}\Big(\cD_\a+(\cD^{\g}\s)\cM_{\g\a}-(\cD_{\a }\s)\cJ\Big)~,
\\
\cDB{}_{\a}~&\to&~\re^{\hf\s}\Big(\cDB_{\a}+(\cDB^{\g}\s){\cM}_{\g\a}
+(\cDB_{\a}\s)\cJ\Big)~,
\\
\cD{}_{a}
~&\to&~\re^{\s}\Big(
\cD_{a}
-\frac{\ri}{2}(\g_a)^{\g\d}(\cD_{(\g}\s)\cDB_{\d)}
-\frac{\ri}{2}(\g_a)^{\g\d}(\cDB_{(\g}\s)\cD_{\d)}
+\ve_{abc}(\cD^b\s)\cM^c
\non\\
&&
+\frac{\ri}{2}(\cD_{\g}\s)(\cDB^{\g}\s)\cM_{a}
-\frac{\ri}{8}(\g_a)^{\g\d}({[}\cD_{\g},\cDB_{\d}{]}\s)\cJ
-\frac{3\ri}{4}(\g_a)^{\g\d}(\cD_{\g}\s)(\cDB_{\d}\s)\cJ
\Big)~.~~~~~~
\eea
\esubeq
The infinitesimal form of this transformation was originally given in \cite{KLT-M11}. 
Making use of \eqref{sW-general}, we can read off 
the super-Weyl transformation of the dimension-one torsion superfields
\bsubeq\label{sW-tor2}
\bea
{\cS}&\to& \frac{1}{4}\re^{\s}
\Big(4{\cS}
+\ri  \cD^\g\cDB_{\g}{\s} \Big)
~,
\\
\cC_{\a \b}&\to&
\Big(
\cC_{\a\b}
-\frac{1}{4}  \big[ \cD_{(\a}, \cDB_{\b)} \big]
\Big)\re^{\s}
~,~~~~~~
\\
\cR&\to&
-\frac{1}{4}
\re^{2\s}\Big(\cDB^2-4\cR\Big)\re^{-\s}
~,
~~~~~~
\bar{\cR} \to
-\frac{1}{4}\re^{2\s}
\Big(\cD^2-4\bar{\cR}\Big)\re^{-\s}
~.
\label{sW-R-2}
\eea
\esubeq

Using the above super-Weyl transformation laws, it is an instructive exercise to demonstrate
that the real vector superfield\footnote{As a consequence of \eqref{N=2-alg-b},
the first term on the right of \eqref{Cotton}
can be written in several equivalent forms: 
$\frac{\ri}{2} \big[\cD^\g ,\bar \cD_\g \big]\cC_{\a\b}=  \ri \cD^\g \bar \cD_\g \cC_{\a\b} = \ri \bar \cD^\g \cD_\g \cC_{\a\b}$.}
\bea
\cW_{\a\b} := \frac{\ri}{2} \big[\cD^\g ,\bar \cD_\g \big]
\cC_{\a\b} - \big[ \cD_{(\a} , \bar \cD_{\b)} \big] \cS 
- 4 \cS \cC_{\a\b}
\label{Cotton}
\eea
transforms homogeneously, 
\bea
\cW_{\a\b} \to \re^{2\s}\, \cW_{\a\b}~.
\label{Cotton-Weyl}
\eea
This superfield is the $\cN=2$ supersymmetric generalization of the Cotton tensor. 
The condition $\cW_{\a\b} =0$,
 which is the equation of motion for $\cN=2$ 
conformal supergravity, is necessary and sufficient for a curved superspace
to be conformally flat. The super-Weyl transformation \eqref{Cotton-Weyl} follows, 
in particular, from the identity
\bea
\frac{\ri}{2} \Big\{ \big[ \cD^\g , \bar \cD_\g \big], \big[ \cD_{(\a} , \bar \cD_{\b)} \big] \Big\} \s &=& 
- 8\cC^{\g}{}_{(\a} 
\cD_{\b)\g}\s 
-4\ri (\cD_{(\a} R) \cD_{\b)} \s  - 4\ri  (\bar \cD_{(\a} \bar R) \bar \cD_{\b)} \s~.~~~
\eea

The super-Weyl and local U$(1)_R$ symmetries can be used to impose the following gauge conditions
\cite{KLT-M11,KT-M11}: $\F_\a = 0$, $\cS=0$ and $\F_{\a\b} = \cC_{\a\b}$. 
In this gauge, which is most suitable to describe the type I supergravity \cite{KT-M11},  
the expression \eqref{Cotton} takes the form $\cW_{\a\b} =\ri \cD^\g \bar \cD_\g \cC_{\a\b} $,
which first appeared in \cite{ZupnikPak}.

We conclude the review of conformal supergravity by recalling the two locally supersymmetric 
and super-Weyl invariant action principles  described in \cite{KLT-M11}.
Given a real scalar Lagrangian $\cL = \bar \cL $ with the super-Weyl transformation law
\bea
\cL \to \re^\s \cL~,
\eea
the functional 
\bea
S&=&\int\rd^3x\rd^2\q\rd^2\qb E \,\cL
~,\qquad
\qquad E^{-1}:= {\rm Ber}(E_A{}^M)
\label{N=2Ac}
\eea
is invariant under the supergravity gauge group. 
Its  super-Weyl invariance follows from the  transformation law of $E$, which is
\bea
E\to \re^{-\s} E~.
\eea
Given a chiral scalar Lagrangian $\cL_{\rm c}$ of U$(1)_R$ charge $-2$ 
and super-Weyl weight two\footnote{Given a covariantly chiral scalar $\F$, ${\bar \cD}_\a \F=0$, 
which transforms homogeneously  under the super-Weyl group, $\F\to \re^{w\s}\F$,  
its super-Weyl weight $w$ 
and its U$(1)_R$ charge have the same value and opposite signs,
that is $[\cJ , \F] = - w \F$, see  \cite{KLT-M11} for more details.}
\bea
\cDB_\a\cL_{\rm c}=0~, \qquad [\cJ, \cL_{\rm c}] =-2 \cL_{\rm c}~, \qquad
\cL_{\rm c}\to\re^{2\s }\cL_{\rm c}~,
\eea
the following {\rm chiral} action
\bea
S_c&=&\int\rd^3x\rd^2\q\rd^2\qb\, \frac{E}{\cR} \,\cL_{\rm c}
=\int\rd^3x\rd^2\q\, \cE \,\cL_{\rm c} 
\label{N=2Ac-chiral}
\eea
is locally supersymmetric and super-Weyl invariant.
Here $\cE$ denotes the chiral density, ${\bar \cD}_\a \cE=0$; 
its explicit form is given by eq. \eqref{5.4}.
The two actions are related to each other 
by the chiral reduction rule 
\bea
\int\rd^3x \rd^2\q \rd^2\bar \q
\, E \,\cL = \int\rd^3x \rd^2\q \rd^2 \bar \q
\, \frac{E}{\cR} \, {\bar \D}\cL
=\int\rd^3x\rd^2\q\, \cE \,\bar{\D}\cL~,
\label{N=2Ac-2}
\eea
where $\bar \D$ denotes the chiral projection operator 
\bea
\bar{\D}:=-\frac{1}{4}(\cDB^2-4\cR)~.
\label{N=2chiralPsi}
\eea
For any scalar  $V$ of U$(1)_R$ weight $q$, 
$\bar \D V$ is a chiral scalar of U$(1)_R$ weight $(q-2)$.\footnote{Covariantly chiral superfields
in  three-dimensional $\cN=2$ supergravity are necessarily scalar under the Lorentz group. 
This is in contrast 
with 4D $\cN=1$ supergravity which allows the existence of covariantly chiral superfields 
with undotted indices, see \cite{BK,GGRS} for reviews.}

For practical calculations, of special importance is 
the following rule for integration by parts  in superspace:
given a vector superfield $V=V^A E_A$,  it holds that 
\bea
\int\rd^3x\,\rd^2\q \rd^2 \bar \q\,  E \,(-1)^{\ve_A}  \cD_AV^A =0~.
\eea

\section{The spinor vielbein}
The supergeometry \eqref{N=2-alg} corresponds to  the conformal supergravity 
constraints \cite{HIPT}.
We now turn to solving these constraints, and expressing all the geometric objects
in \eqref{CovDev}, in terms of unconstrained prepotentials. 

It follows from \eqref{N=2-alg-a} that $T_{\a\b}{}^C = 0 $. Then eq. \eqref{2.4a}
tells us that 
\bea
\{E_\a , E_\b \} = C_{\a\b}{}^\g E_\g ~. 
\eea 
In accordance with the Frobenius theorem (see, e.g. \cite{Warner}), this is solved by 
\bea
E_\a = F N_\a{}^\m \UE_\m~, \qquad  N= (N_\a{}^\m) \in {\rm SL}(2,{\mathbb C})~,
\label{spinor_vilebein}
\eea
where $F N_\a{}^\m$ is a nonsingular $2\times 2$ matrix, and the first-order operators 
$\UE_\m$ are such that 
\bea
\{ \UE_\m , \UE_\n\} =0~.
\label{3.3}
\eea
A general solution of \eqref{3.3} is 
\bea
\UE_\m = \re^W D_\m \re^{-W} ~, \qquad W = W^M D_M ~, 
\eea
with $D_M = (\pa_m , D_\m , \bar D^\m)$ the flat superspace covariant derivatives.
The prepotential $W^M$ is complex, $\bar W \neq W$. It 
 is defined modulo arbitrary gauge transformations 
\bea
\re^W \to \re^W \re^{-\bar \L} ~,
\label{labmda-tran}
\eea 
where $\bar \L $ denotes the complex conjugate of a constrained vector field
\begin{subequations} \label{LLL}
\bea
\L = \L^m \pa_m+  \L^\m  D_\m +  \bar \r_\m \bar D^\m ~, \qquad 
[\bar D_\m , \L ] \propto \bar D_\n~.
\eea
The constraints on $\L$ are solved by 
\bea
\L^{\m\n} :=  \L^m(\g_m)^{\m\n}  = -{\ri} ( \bar D^\m L^\n +  \bar D^\n L^\m)~,
\qquad \L^\m = -\frac{1}{4} \bar D^2 L^\m~,
 \label{LLL-b}
\eea
\end{subequations}
with $\bar \r_\m$ and $L^\m$ unconstrained spinor superfields.  
The transformation \eqref{labmda-tran} should leave the spinor vielbein, $E_\a$, unchanged, and therefore $F$ and $N_\a{}^\m$ should transform in a certain way. 
In the infinitesimal case, their transformation laws are as follows:
\begin{subequations}
\bea
\d F &=&\hf F \re^W D^\m \r_\m ~, \\
\d N_\a{}^\m &=& -N_\a{}^\n \re^W D_{(\n} \r^{\m)} ~.
\eea
\end{subequations}

Under a general coordinate transformation generated by $K^C$,  eq. \eqref{tau},
 the prepotential $W$ changes by a left shift
\bea
\re^W \to \re^K \re^W~, \qquad K = K^C D_C = \bar K~.
\eea 

The above discussion is almost identical to Siegel's analysis in four dimensions \cite{Siegel}.
Now comes a new feature of the 3D case.  
Consider the {\it complex} unimodular matrix $N =(N_\a{}^\m)$ in \eqref{spinor_vilebein}.
Under a local Lorentz transformation described by the parameter $l^{\g\d}$ in \eqref{tau}, 
the matrix $N$ transforms as 
\bea
N_\a{}^\m \to (\re^l )_\a{}^\b N_\b{}^\m~, \qquad \re^l \in {\rm SL}(2,{\mathbb R})~.
\eea
In the four-dimensional case, the Lorentz group was ${\rm SL}(2,{\mathbb C})$, 
and the local Lorentz freedom was sufficient to gauge away $N$, for instance  to choose
a gauge $N = {\mathbbm 1}$. This is no longer the case in three dimensions, 
for we can not gauge away a part of $N$ parametrizing the right coset space 
$ {\rm SL} (2,{\mathbb R}) \backslash {\rm SL} (2,{\mathbb C})$.

The following unimodular matrix
\bea
{\frak J}= ( {\frak J}_\m{}^\n ) := N^{-1} \bar N \in {\rm SL}(2,{\mathbb C})
\label{jay}
\eea
is Lorentz invariant.  Its main property is 
\bea
{\frak J} \bar {\frak J} = {\mathbbm 1}~,
\eea
which is solved by 
\begin{align} 
{\frak J}  = \left(\begin{array}{cc}
a & \ri \b \\
\ri \g  & \bar a
\end{array}\right)~, \qquad |a|^2 + \b \g =1~, \qquad 
\a \in {\mathbb C}~, \qquad \b, \g \in {\mathbb R} ~.
\end{align}
We are going to show that ${\frak J}$ is uniquely expressed in terms of the prepotential $W^M$.

Let us introduce a semi-covariant vielbein of the form
\begin{subequations} \label{SCV}
\bea
\UE_M = (\UE_m , \UE_\m , \bar \UE^\m ) = \UE_M{}^ND_N~,
\qquad \UE^{-1} := {\rm Ber}(\UE_M{}^N) ~.
 \label{SCV-a}
\eea
where $\UE_{\m\n} =\UE_{\n\m} := \UE_m(\g^m)_{\m \n}  $ is defined by 
\bea
\{ \UE_\m , \bar \UE_\n \} = -2\ri \UE_{\m\n} - 2  \ve_{\m \n} \X ~,\qquad
\X= \X^N\UE_N
=\X^n \UE_n +\X^\n \UE_\n +\bar \X_\n \bar \UE^\n = \bar \X~.~~~~~~
 \label{SCV-b}
\eea
\end{subequations}
It follows from the  relation \eqref{N=2-alg-b} that the {\it vectorial} part of 
the first-order operator
$N_\a{}^\m \bar N_\b{}^\n \{ \UE_\m , \bar \UE_\n \} $
must be symmetric in $\a$ and $\b$,
\bea
N_\a{}^\m \bar N_\b{}^\n \{ \UE_\m , \bar \UE_\n \} 
= N_\b{}^\m \bar N_\a{}^\n \{ \UE_\m , \bar \UE_\n \} +\cdots ~,
\label{3.15}
\eea
where the ellipsis denotes all terms with spinor derivatives 
$ \UE_\r $ and $\bar \UE_\r $.
One may see that the condition \eqref{3.15} is equivalent to 
\bea
{\frak J}^{(\m\n)} = \frac{\ri}{2} \tr {\frak J} \, \X^{\m\n}~.
\eea
By construction, ${\frak J}^{\m\n}  := \ve^{\m\l}{\frak J}_\l{}^\n
= {\frak J}^{(\m\n)} +\hf \ve^{\m\n} \tr {\frak J}$.
Now, using the  condition that $\frak J$ is unimodular, 
\bea
{\frak J}^{\m\l} {\frak J}^{\n \r} \ve_{\l \r} = - \ve^{\m\n}~,
\eea
we derive the relation 
\bea
\Big( \hf \tr {\frak J} \Big)^2\Big\{ 1+ \X^m \X_m\Big\} =1~.
\eea
This  leads to the final expression for $\frak J$:
\bea
{\frak J}_\m{}^\n = \frac{ \d_\m{}^\n +\ri \, \X_\m{}^\n}{ \sqrt{1 + \X^2}}~, 
\qquad \X^2 :=  \X^m \X_m~.
\label{3.19}
\eea
We have chosen the sign of the square root so that  
${\frak J}_\m{}^\n =  \d_\m{}^\n $ when $W^M$ and thus $ \X^m $ vanish. 
The matrix $\frak J$ is completely determined as a power series in $W^M$ and its derivatives.  
The solution \eqref{3.19}  becomes singular at $\X^2 =-1$.
This is analogous to the situation in (2,2) supergravity in two dimensions
\cite{GrisaruW1,GrisaruW2}.

\section{The spinor connection and the torsion superfields}  

In accordance with  \eqref{N=2-alg-a}, the U$(1)_R$ curvature $F_{\a\b}$ is equal to zero, 
and therefore
\begin{subequations}
\bea
\cD_\a &=& E_\a +\O_\a{}^{\g\d } \cM_{\g\d} - E_\a U \cJ \quad \longleftrightarrow
\quad \F_\a =\phantom{-} \ri E_\a U~, 
 \\
\bar \cD_\a &=& \bar E_\a + {\bar \O}_\a{}^{\g\d } \cM_{\g\d} + \bar E_\a \bar U \cJ
\quad \longleftrightarrow
\quad \bar \F_\a = -\ri \bar E_\a \bar U~, 
\eea
\end{subequations}
for some complex scalar prepotential $U$ which is defined modulo gauge transformations
of the form 
\bea
U \to U +2\bar \o~, \qquad \cD_\a \bar \o=0~, \qquad [\cJ, \bar \o] =0~.
\label{lambda}
\eea
Under the  super-Weyl 
and local U(1)${}_R$ transformations, this prepotential can be seen to change as follows:
\begin{subequations}\label{Weyl+chiral}
\bea
U &\to& U+\s~, \ \\
U &\to&U+ \ri \, \t~. 
\eea
\end{subequations}
For the torsion $\cS$, it is a trivial exercise to deduce from \eqref{N=2-alg-b} that 
\bea
{\cS} = \frac{\ri}{4}  \cD^\a\bar \cD_\a \cV~, \qquad
\cV=\hf (U +\bar U)~.
\eea
These relations have already been given in \cite{KT-M11}.
The $\o$-gauge transformation of  $\cV$ is 
\bea
\cV \to \cV+ \o + \bar \o~.
\label{lambda-S}
\eea
It is natural to interpret $\cV$ as the gauge prepotential of an Abelian  vector supermultiplet, 
and the torsion $\cS$ as the corresponding gauge invariant field strength. 

Consider a covariantly chiral superfield $\J$ of U$(1)_R$ charge $q$, 
\bea
\bar \cD_\a \J =0~, \qquad [\cJ, \J] = q \J~.
\eea
It can be represented in the form 
\bea
\J = \re^{-q \bar U} \re^{\bar W} \hat{\J} ~, \qquad \bar D_\a \hat{\J} =0~.
\label{J-definition}
\eea
Here $\hat \J$ is an arbitrary {\it flat} chiral superfield.
The representation in which the covariantly chiral superfield $\J$ is described by 
$\hat \J$ is called {\it chiral}.

The spinor Lorentz connection is not an independent field as a consequence of the constraint
\bea
0= T_{\a\b}{}^\g = C_{\a\b}{}^\g + \O_{\a\b}{}^\g + \O_{\b\a}{}^\g 
+\ri (\F_\a \d_\b{}^\g + \F_\b \d_\a{}^\g) ~.
\eea
It allows us to uniquely determine the spinor Lorentz connection
\bea
\O_{\a\b\g} = \hf \Big( C_{\b\g\a} - C_{\a\b\g} - C_{\g\a\b} \Big) 
+\ri \Big(\ve_{\a\b} \F_\g + \ve_{\a\g} \F_\b \Big)~.
\eea
This involves the  anholonomy coefficients
\bea
C_{\a\b}{}^\g = E_\a \ln F \d_\b{}^\g + E_\b \ln F  \d_\a{}^\g 
+\Big(E_\a N_\b{}^\m +E_\b N_\a{}^\m\Big) (N^{-1})_\m{}^\g ~.
\eea

To compute the  antichiral scalar $\bar \cR$, it is convenient to perform a 
{\it complex} Lorentz transformation which results in 
\begin{subequations} \label{LG}
\bea
N=\mathbbm 1~,  \qquad \bar N  = {\frak J} \neq {\mathbbm 1}~,
\label{LG_a}
\eea
and hence 
\bea
E_\a = F\d_\a{}^\m \UE_\m ~.
\label{LG_b}
\eea 
\end{subequations}
In such a gauge, it is a short calculation to compute 
the torsion $\bar \cR$. The result is 
\begin{subequations}
\bea
\bar \cR = -\frac{1}{4} \re^{2U} \UE^2 \big(F^2 \re^{-2U}\big)~.
\label{barR}
\eea
Since the right-hand side is invariant under arbitrary Lorentz transformations, 
the relation \eqref{barR} holds in general. Taking the complex conjugate of \eqref{barR} gives
\bea
\cR = -\frac{1}{4} \re^{2\bar U} \bar \UE^2 \big(\bar F^2 \re^{-2\bar U}\big)~.
\label{R-definition}
\eea
\end{subequations}
These results have a simple generalization: 
given a scalar superfield $\U$ of U$(1)_R$ charge $q$, 
\bea
[\cJ ,\U ]= q \U~,
\eea
one may show that 
\begin{subequations}
\bea
\big( \cD^2- 4\bar \cR\big) \U &=& \re^{(2+q)U} \UE^2 \Big( F^2 \re^{-(2+q)U}\U\Big)~, \\
\big(\bar \cD^2- 4 \cR\big) \U &=& 
\re^{(2-q)\bar U} \bar \UE^2 \Big( \bar F^2 \re^{-(2-q)\bar U} \U\Big)~.
\eea
\end{subequations}

Our next task is to compute $E^{-1} ={\rm Ber} (E_A{}^M)$ and $F$ in terms of prepotentials.
For this, we will obtain a useful expression for the spinor superfield 
\bea
T_\a:= (-1)^{\e_B} T_{\a B}{}^B ~,
\eea
 which, in accordance with  \eqref{N=2-alg}, 
is equal to zero, $T_\a =0$.\footnote{The dimension of $T_\a$ is one-half, but all components  
of the torsion have dimension greater than or equal to one.}
On the other hand, if one explicitly evaluates $T_\a$, 
the condition $T_\a=0$ proves to contain nontrivial information. From \eqref{2.4a} we deduce
\bea
(-1)^{\e_B} T_{AB}{}^B = (-1)^{\e_B} C_{AB}{}^B - \O_{BA}{}^B -\ri q_A \F_A ~,
\eea
and thus 
\bea
T_\a = (-1)^{\e_B} C_{\a B}{}^B - \O_{\b \a}{}^\b -\ri  \F_\a~.
\eea 
Further, it may be shown (see, e.g., \cite{BK,GGRS} for derivations) that 
\bea
(-1)^{\e_B} C_{\a B}{}^B = -E_\a \ln E - (1\cdot {\stackrel{\leftarrow}{E}_\a}) ~.
\eea
Since $T_\a$ is a spinor superfield, 
we can evaluate it  by choosing a useful Lorentz gauge, and then restore the general answer 
simply by performing the inverse Lorentz transformation. Such a Lorentz gauge condition is 
defined in \eqref{LG}.  In this gauge we obtain 
$\O_{\b\a}{}^\b = - 3\d_\a{}^\m \re^U \UE_\m (F\re^{-U} )$. 
Putting together the building blocks obtained gives 
\bea
T_\a = E_\a \Big[E^{-1} F^2 \re^{-2U} (1\cdot \re^{\stackrel{\leftarrow}{W} }) \Big] ~.
\eea
Setting $T_\a =0$ gives the following fundamental result 
\bea
E^{-1} F^2 \re^{-2U} (1\cdot \re^{\stackrel{\leftarrow}{W} }) = \bar \vf^{-4}~, \qquad
E_\a \bar \vf =0~, 
\label{vf-definition}
\eea
for some antichiral superfield $\bar \vf$.
One may see that the chiral superfield $ \vf$ 
is invariant under the super-Weyl and local U$(1)_R$ 
transformations. It can be represented as 
\bea
\vf =  \re^{\bar W} \hat{\vf} ~, \qquad \bar D_\a \hat{\vf} =0~.
\label{vf-flat}
\eea
Under the gauge transformation \eqref{lambda}, its 
transformation law is 
\bea
\vf ~\to ~ \re^{ \o } \vf~.
\label{lambda-vf}
\eea
The gauge freedom \eqref{lambda} can be used to completely gauge away $\vf$, 
for instance to choose a gauge $\vf =1$.

Another important result follows from the relations \eqref{N=2-alg-b}, 
\eqref{spinor_vilebein} and
\eqref{SCV}
\bea
E = (F \bar F)^{-1}  \UE ~.
\label{4.23}
\eea
The equations  \eqref{vf-definition} and \eqref{4.23} lead to 
\bea
E = \bar \vf  \re^{-\cV}\vf \,  \Big\{ \UE^2 (1\cdot \re^{\stackrel{\leftarrow}{W} })
(1\cdot \re^{\stackrel{\leftarrow}{\bar W} }) \Big\}^{1/4}
\eea
It is seen that $E$ is invariant under the gauge transformation \eqref{lambda-S}
and \eqref{lambda-vf}, as it should be of course. 
Finally, we derive the explicit expression for $F$:
\bea
F =  \left\{ \UE^2  \frac{ \vf^4 \re^{-2\bar U} }{ (\bar \vf^{4} \re^{-2U} )^3} \,
\frac{ (1\cdot \re^{\stackrel{\leftarrow}{\bar W} })} { (1\cdot \re^{\stackrel{\leftarrow}{W} })^3}
\right\}^{1/8}~.
\eea

The building blocks constructed are sufficient to read off explicit expressions 
for the vector covariant derivative $\cD_a$ and the torsion $\cC_{\a\b}$, simply 
by making use of  \eqref{N=2-alg-b}. 
The latter is also the Lorentz curvature
\bea
\cC^{\g\d} = \frac{1}{4} \ve^{\a\b} R_{\a\beu}{}^{\g\d}~.
\eea
This can be evaluated using  eq.  \eqref{2.4b} and the anholonomy coefficients 
appearing in 
\bea
\{E_\a , \bar E_\b  \} =  C_{\a \beu}{}^C E_C = C_{\a \beu}{}^c E_c 
+ C_{\a \beu}{}^\g E_\g  + C_{\a \beu\, \gau}\bar E^\g~.
\label{4.27}
\eea
The anholonomy coefficients are uniquely determined in terms of the spinor connections.
Indeed, it follows from \eqref{2.4a} and \eqref{N=2-alg-b} that 
\begin{subequations} \label{4.28}
\bea
T_{\a \beu}{}^c &=& C_{\a \beu}{}^c = -2\ri (\g^c)_{\a\b}~, \\
T_{\a \beu}{}^\g &=& C_{\a \beu}{}^\g +  \bar \O_{ \b \a}{}^\g 
+ \bar \UE_\b \bar U \d_\a{}^\g =0~,\\ 
T_{\a \beu}{}^{\gau} &=& C_{\a \beu}{}^{\gau} +  \O_{\a \b}{}^\g 
+  \UE_\a  U \d_\b{}^\g =0~.
\eea
\end{subequations}
We thus obtain 
\bea
\cC^{\g\d} = \frac{1}{4} \ve^{\a\b} \Big\{ E_\a \bar \O_\b{}^{\g\d} + \bar E_\b \O_\a{}^{\g\d}
&-& C_{\a \beu}{}^\l \O_\l{}^{\g\d} + C_{\a \beu}{}^{\underline \l} \bar \O_\l{}^{\g\d} \non \\
&+& \O_\a{}^{\g \l} \bar \O_{\b\l}{}^\d + \O_\a{}^{\d\l} \bar \O_{\b \l}{}^\g \Big\}~.
\eea
Using eqs. \eqref{4.27} and \eqref{4.28} we can read off the vector vielbein $E_a$.

\section{Chiral action}
The chiral density $\cE$, which determines the chiral action principle \eqref{N=2Ac-chiral},
remains the only geometric quantity which we have not yet  expressed in terms of prepotentials.
The present section is aimed at filling this gap. 

In accordance with  \eqref{J-definition}, the chiral  Lagrangian can be written as
\bea
\cL_{\rm c} =  \re^{2 \bar U} \re^{\bar W} \hat{\cL}_{\rm c}~, 
\qquad \bar D_\a \hat{\cL}_{\rm c}=0~.
\eea 
We also recall the explicit form of $\vf$ given by eq. \eqref{vf-flat}.
Using the representations \eqref{R-definition} and  \eqref{vf-definition}, we then obtain 
\bea
 \frac{ E}{\cR}\, \cL_{\rm c}  = 
-4  \frac{\bar F^2 \re^{-2\bar U}}
 { \bar \UE^2 \big(\bar F^2 \re^{-2\bar U}\big)} 
 (1\cdot \re^{\stackrel{\leftarrow}{\bar W} }) \re^{\bar W} ( \hat{\vf}^4 \hat{\cL}_{\rm c})~.
 \label{5.2}
 \eea
Next, we should recall an important theorem \cite{GS80} (see \cite{BK} for a detailed proof).

{\it Theorem}.
Given a change of superspace variables of the form $z^M \to \hat{z}{}^M =  \re^{\bar W} z^M$,
its Jacobian is equal to  $ (1\cdot \re^{\stackrel{\leftarrow}{\bar W} })$, 
\bea
\hat{z}{}^M =  \re^{\bar W} z^M \quad \Longrightarrow \quad 
{\rm Ber} (\pa_M \hat{z}^N ) =  (1\cdot \re^{\stackrel{\leftarrow}{\bar W} })~.
\eea

Along with the theorem stated, it only remains to note that 
the operator $\bar \UE^\a $ appearing in \eqref{5.2} becomes a partial derivatives 
in the new coordinate system  introduced,
$\bar \UE^\a = \pa/ \pa\hat{\bar \q}_\a$. 
Now, in the chiral action \eqref{N=2Ac-chiral} we make use of \eqref{5.2},
perform the change of variable $z^M \to \hat{z}{}^M =  \re^{\bar W} z^M$, 
and then carry out the integral over the Grassmann variables $\hat{\bar \q}_\a$.
This leads to 
\bea
\int {\rm d}^3x {\rm d}^2 \q \rd^2 \bar \q \, \frac{ E}{\cR}\, \cL_{\rm c} 
= \int {\rm d}^3 \hat{x} {\rm d}^2 \hat{\q} \,\hat{\vf}^4 \hat{\cL}_{\rm c}~.
\label{5.4}
\eea
We conclude that, in the chiral representation, the chiral density 
  ${\hat \cE}$ coincides with $  {\hat \vf}^4$.

\section{Discussion}

We have expressed the geometric potentials $E_A{}^M$, $\O_A{}^{\g\d} $ and $\F_A$ 
in terms of several complex prepotentials (and their complex conjugates):
(i) the supervector $W^M$; (ii) the unimodular matrix $N=(N_\a{}^\m) \in {\rm SL}(2,{\mathbb C})$; 
(iii) the scalar $U$; and (iv) the chiral superfield $\vf$. 
Now, we are in a position to demonstrate that the gauge freedom available in the theory 
allows us to algebraically gauge away all prepotentials except
 a real vector $H^m =\bar H^m$ that may be identified with the gravitational  superfield. 
First of all, the local Lorentz invariance associated with the parameter $l^{ \g\d }$ in \eqref{tau} 
can be fixed by gauging away three of the six {\it real} degrees of freedom encoded in  $N$,  
leaving only the Lorentz-invariant matrix $\frak J$ defined by \eqref{jay}.
The latter is uniquely expressed in terms of $W^M$ and its conjugate $\bar W^M$, according to 
eq. \eqref{3.19}. Secondly, it follows from \eqref{Weyl+chiral} that the scalar superfield $U $
can be gauged away by applying a super-Weyl and local U$(1)_R$ transformation, 
which completely fixes these two gauge symmetries. Thirdly, it follows from 
\eqref{lambda-vf} that the $\o$-gauge freedom can be used to gauge away $\vf$.
As a result, we stay only with the prepotential $W^M$ with the gauge transformation law
\bea
\re^W \to \re^K \re^W \re^{-\bar \L}~,
\label{6.1}
\eea
where the parameters $K=K^CD_C = K$ and $\L=\L^M D_M$ correspond to the general coordinate 
transformations \eqref{tau} and $\L$-transformations  \eqref{LLL} respectively.
The general coordinate invariance alone can be used to gauge away the real part of  $W$
by choosing
\bea
W= -\ri H~, \qquad   H=H^M D_M = H~.
\eea 
In conjunction with the $\L$-gauge freedom, we can further gauge away 
the spinor components of $H^M$ thus arriving at the gauge  
condition\footnote{There exists an alternative gauge  condition: 
$\re^{-\bar W} x^m = x^m + \ri \cH^m (x, \q, \bar \q)$, $\re^{-\bar W} \q^\m = \q^\m $
and $\re^{-\bar W} \bar \q_\m = \bar \q_\m $. Here the vector superfield 
$\cH^m$ is real, $\bar \cH^m  = \cH^m$.
Using this gauge condition, one may develop a 3D version 
of the geometric  supergravity formalism due to Ogievetsky and Sokatchev
\cite{OS2}, see \cite{BK} for a review. Such a generalization was sketched 
by Zupnik and Pak long ago \cite{ZupnikPak}. In spite of its nice geometric structure, 
the Ogievetsky-Sokatchev has not found applications for quantum calculations 
(unlike the approach advocated in \cite{GGRS})  
and we do not elaborate on it here; see however \cite{ZupnikPak2}.}
\bea
W = {-\ri H} ~, \qquad H=  \bar H = H^m \pa_m ~.
\label{vectorH}
\eea
As in four dimensions \cite{Siegel,SG},
the general coordinate group is automatically eliminated if one works with the operator
\bea
\re^{-2 \ri H} := \re^{-\bar W} \re^W~,
\eea
which is invariant under the $K$-transformations \eqref{6.1}.
The infinitesimal transformation law  of $H$ is 
\bea
\d \re^{-2 \ri H}   =  \L  \re^{-2\ri H}  - \re^{-2\ri H}  \bar \L~.
\label{6.4} 
\eea
To preserve the gauge condition \eqref{vectorH}, the spinor parameter $\bar \r_\m$ 
in \eqref{LLL} has to be be constrained as
\bea
\bar \r_\m = \re^{-2\ri H} \bar \L_\m~.
\label{6.5}
\eea
This follows from   \eqref{6.4} by requiring that all terms with spinor derivatives in the right-hand
side cancel out. Making use of eqs. \eqref{LLL-b} and \eqref{6.5}, it may be seen 
that the transformation law \eqref{6.4} can indeed be cast in the form \eqref{1.1}.

So far we have only discussed the prepotential structure of $\cN=2$ conformal supergravity 
in three dimensions. The analysis can easily be extended to the the off-shell 
versions of $\cN=2$ Poincar\'e and anti-de Sitter supergravity theories 
constructed in \cite{KT-M11}.  This is achieved by coupling the Weyl supermultiplet to an appropriate conformal compensator in a super-Weyl invariant way.
There are two minimal Poincar\'e supergravity theories with $8+8$ 
off-shell degrees of freedom, called type I and type II theories in \cite{KT-M11}. Their extensions
with a cosmological term, which were constructed in \cite{KT-M11}, 
are also known as  (1,1) and (2,0) AdS supergravity theories, following the terminology of \cite{AT}.
The type I theory is a 3D generalization of the old minimal formulation for 
$\cN=1$ supergravity in four dimensions \cite{WZ,old}. 
It makes use of a covariantly chiral scalar $\F$ of U$(1)_R$ weight $-1/2$ and super-Weyl weight 
$+1/2$,
\bea
\cDB_\a \F=0~, \qquad [\cJ, \F] =-\hf \F~, \qquad
\F \to\re^{\hf\s }\F~.
\eea
The super-Weyl and local U$(1)_R$ gauge symmetries 
may be fixed by imposing the gauge condition $\F=1$.
The type II supergravity is a 3D generalization of the new minimal formulation for 
$\cN=1$ supergravity in four dimensions constructed at the linearized level in \cite{AVS}
and later at the full nonlinear level in \cite{new}.
Its conformal compensator is a vector multiplet  described by  a real scalar prepotential
$G$ which is defined modulo arbitrary gauge transformations of the form
\be 
\d G = \l + \bar \l~, \qquad {\bar \cD}_\a \l =0~,\qquad  [\cJ , \l] =0~, 
\ee
and is inert under the super-Weyl transformations.
Associated with $G$ is the gauge-invariant field strength 
\bea
\mathbb G= \ri {\bar \cD}^\a \cD_\a G = \bar {\mathbb G}~,
\label{G-prep}
\eea
which is covariantly linear,
\bea
(\cD^2 - 4 \bar R ) \mathbb G =  
(\cDB^2-4R)\mathbb G= 0~.
\label{N=2realLinear}
\eea
The field strength is required to be nowhere vanishing, $\mathbb G \neq 0$.
The super-Weyl transformation law of $\mathbb G$ proves to be 
\bea
{\mathbb G} \to\re^{\s }{\mathbb G}~.
\eea
The super-Weyl gauge freedom may be fixed by imposing the gauge condition 
${\mathbb G}=1$.

Similar to the four-dimensional case \cite{SG,non-min}, 
there exists a non-minimal formulation for 3D $\cN=2$ Poincar\'e supergravity. 
It makes use of a complex linear compensator $\S$ 
of U$(1)_R$ weight $(1-w)$ and super-Weyl weight 
$w$,
\bea
(\cDB^2-4R)\S=0~, \qquad [\cJ, \S] = (1-w)\S~, \qquad
\S \to\re^{w\s }\S~,
\eea
with $w$ a real parameter which specifies the off-shell supergravity formulation 
under consideration.\footnote{The parameter $w$ is related to the 4D non-minimal parameter $n$
\cite{SG} as follows $w = (1-n)/(1+3n)$.} 
The compensator has to be  nowhere vanishing, $\S \neq 0$.

In four dimensions, it was long believed that $\cN=1$ AdS supergravity could not be described
using a non-minimal set of auxiliary fields \cite{GGRS}. 
Nevertheless, such a formulation  has  recently been  constructed 
\cite{BKdual} in the case $n=-1$  by using a deformed complex linear constraint obeyed by the compensator. This construction has also been extended to the 3D $\cN=2$ case \cite{KT-M11}.
In three dimensions, non-minimal AdS supergravity can consistently be defined 
by choosing  $w=-1$ and  considering a deformed complex linear compensator $\G$
constrained by 
 \begin{align}
-\frac{1}{4} (\bar \cD^2 - 4 R) \Gamma = \m ={\rm const}~.
\end{align}
The complex parameter $\m$ is related to the cosmological constant, see \cite{KT-M11} 
for more details. The resulting non-minimal formulation describes the (1,1) AdS supergravity.
\\

\noindent
{\bf Acknowledgements:} \\
The author is grateful to Joseph Novak for reading the manuscript.
This work is supported in part by the Australian Research Council, 
project No. DP1096372.

\begin{footnotesize}

\end{footnotesize}

\end{document}